\documentclass[conference,compsoc]{IEEEtran}
\usepackage{filecontents}
\usepackage{stfloats}
\usepackage{placeins}
\usepackage{nameref}
\usepackage{hyperref}
\usepackage{mdframed}

\usepackage{epsf}
\usepackage{epsfig}
\usepackage{graphicx}
\usepackage{times}
\usepackage{ifthen}
\usepackage{array}
\usepackage{comment}
\usepackage{url}
\usepackage{cellspace}
\usepackage{latexsym}
\usepackage{amsfonts}
\usepackage{subfigure}
\usepackage{multirow}
\usepackage{multicol}
\usepackage{tabularx}
\usepackage{listings}
\usepackage{amssymb}
\usepackage{mathtools}
\usepackage{wrapfig}
\usepackage{xspace}
\usepackage{enumitem}
\usepackage{footnote}
\usepackage{gensymb}
\usepackage{todonotes}
\usepackage{tabularx}
\usepackage{makecell}
\usepackage{xcolor}
\newboolean{showcomments}
\usepackage{tcolorbox}

\usepackage[utf8]{inputenc}
\usepackage{tikz}

\setboolean{showcomments}{true}

\ifthenelse{\boolean{showcomments}}
  {\newcommand{\nb}[2]{
    \fbox{\bfseries\sffamily\scriptsize#1}
    {\sf\small$\blacktriangleright$\textit{#2}$\blacktriangleleft$}
   }
   
  }
  {\newcommand{\nb}[2]{}
   
  }

\newcommand\TODO[1]{{\color{red} \nb{TODO}{#1}}}

\newcommand{\gpt}{\texttt {GPT}\xspace}
\newcommand{\gemini}{\texttt {Gemini}\xspace}

\newcommand{\ie}{\textit{i.e.},\xspace}
\newcommand{\eg}{\textit{e.g.},\xspace}

\newcommand{\etal}{\textit{et al.}\xspace}

\newcommand{\promptnum}[1]{\texttt{\small Prompt}\textsubscript{#1}}

\newcommand{\fnumber}[1]{{$\mathcal{F}_{#1}$}}
\newcommand\dgnumber[1]{{$\mathcal{DG}_{#1}$}}
\newcommand\funcnumber[1]{\colorbox{gray!20}{{$\mathcal{FN}_{#1}$}}}
\newcommand{\funcnumbers}{\colorbox{gray!20}{{$\mathcal{FN}_{S}\xspace$}}}

\newcommand{\resq}[1]{\textbf{$RQ_{#1}$}}
\newcommand{\finding}[2]{
\begin{tcolorbox}[colback=white, colframe=black, boxrule=0.3pt, left=3pt, right=3pt, top=3pt, bottom=3pt]
\noindent{\bf {\em Finding\ {#1}}} (\fnumber{{#1}}): {#2}.
\end{tcolorbox}
}

\newcommand{\secref}[1]{Sec.~\ref{#1}\xspace}
\newcommand{\figref}[1]{Fig.~\ref{#1}\xspace}
\newcommand{\tabref}[1]{Table~\ref{#1}\xspace}

\newcounter{taskctr}

\newcommand{\myparagraph}[1]{\vspace{0.5em}\noindent{\bf #1}:}

\definecolor{dkgreen}{rgb}{0,0.6,0}
\definecolor{gray}{rgb}{0.5,0.5,0.5}
\definecolor{mauve}{rgb}{0.58,0,0.82}
\definecolor{shadecolor}{rgb}{0.95,0.95,0.95}
\setlist[itemize]{leftmargin=*, noitemsep, topsep=1pt}
\setlist[enumerate]{leftmargin=*, noitemsep, topsep=1pt}

\makeatletter
\renewcommand{\paragraph}{%
	\@startsection{paragraph}{4}%
	{\z@}{0.5ex \@plus 0ex \@minus .2ex}{-1em}%
	{\normalfont\normalsize\bfseries}%
}
\makeatother

\makeatletter
\g@addto@macro\normalsize{%
	\setlength\abovedisplayskip{0pt}
	\setlength\belowdisplayskip{0pt}
	\setlength\abovedisplayshortskip{-10pt}
	\setlength\belowdisplayshortskip{0pt}
}
\makeatother

\DeclareGraphicsExtensions{.pdf,.png}
\graphicspath{{img/}}

\newcolumntype{P}[1]{>{\centering\arraybackslash}p{#1}}

\renewcommand{\refname}{{\hskip 0pt plus 0.1fil minus 0pt} References Cited\hfil}                   

\begin{document}

\date{}

\title{\Large \bf A Systematic Evaluation of Traditional Privacy Policy Analysis Tools Against LLMs}

\author{
    \IEEEauthorblockN{
        Madhav Aryal\IEEEauthorrefmark{1},
        Sudipa Saha\IEEEauthorrefmark{1},
        Sunil Manandhar\IEEEauthorrefmark{2},
        Anshuman Chhabra\IEEEauthorrefmark{1}, and
        Kaushal Kafle\IEEEauthorrefmark{1}
    }\\[0.5ex]
    \IEEEauthorblockA{\IEEEauthorrefmark{1}University of South Florida, \{madhavaryal, saha134, anshumanc, kafle\}@usf.edu} \\
    \IEEEauthorblockA{\IEEEauthorrefmark{2}IBM T.J. Watson Research Center, sunil@ibm.com}
}

\maketitle

\begin{abstract}
The advent of LLMs has significantly changed the research on privacy policy and data compliance analysis by enabling tasks that previously required specialized, domain-specific tools. However, it remains unclear to what extent LLMs can truly replicate the diverse functionalities, and the wide range of methodologies and analysis offered by prior work.
In this paper, we conduct the first systematic evaluation of whether off-the-shelf LLMs can replace specialized privacy analysis tools. 
We study six representative tools spanning three major functionalities: contradiction detection, regulatory compliance analysis, and privacy policy summarization and aggregation, and across three intermediate tasks: structured data extraction using tuples, Semantic Role Labeling (SRL) and manual privacy policy labeling. 
We compare the performance of two state-of-the-art LLMs (GPT-5.2 and Gemini-2.5 in various configurations) against the tools by directly prompting the models to perform corresponding functionalities and tasks on a custom dataset of 10 privacy policies, allowing us to assess whether off-the-shelf models can produce tool-specific functionalities without further engineering or domain-specific training, major limitations in prior work. 
Our results show that LLMs consistently match or exceed the capabilities of existing tools across the functionalities. For instance, LLMs identified substantially more intra-policy contradictions than prior work, outperformed the tools in regulatory analysis tasks, generated more comprehensive results in privacy policy aggregation tasks and even outperformed traditional tools in intermediate tasks. 
In manual labeling of first-party collection entities, LLMs achieved an average precision of 81.8\% and recall of 70.9\%, while for labeling of third-party sharing entities, they achieved an average precision of 91.4\% and recall of 70.8\% compared to the OPP-115 dataset. 
Overall, our findings indicate that LLMs can effectively perform a broad range of functionalities and tasks in privacy policy and regulation analysis that previously required specialized tools. We further discuss the implications of our findings for future privacy research.

\end{abstract}

\section{Introduction}
As data privacy concerns grow, there have been an increasing number of data protection regulations introduced all over the world (\eg GDPR in EU, PIPEDA in Canada, PIPL in China), and in many US states (\eg CPRA~\cite{cpra} in California, TDPSA~\cite{tdpsa} in Texas, FDBR~\cite{fdbr} in Florida). 
To comply with these regulations, organizations that collect, process and share user data often are {\em required to} provide their data privacy disclosures in the form of an easily accessible privacy policy. 
Hence, privacy policy analysis has been a prominent research area in the academic community to help readers understand and interpret these privacy policies. 
To that end, prior work has produced a diverse ecosystem of tools to address different aspects of this problem, including the detection of inconsistencies within policy statements~\cite{Andow:SECURITY2019}, conformity with regulations\cite{manandhar-sec22}, mapping policy statements to app behavior~\cite{Andow:SECURITY2020}, and policy interpretation and summarization~\cite{Harkous:SECURITY2018}. 
To achieve this, prior work relied primarily on Natural Language Processing (NLP) techniques to create machine learning (ML) models that targeted specific use cases for a particular regulation (\eg violations of specific GDPR articles).

Recently, the emergence of Large Language Models (LLMs) has fundamentally changed the landscape of NLP-based analysis. 
LLMs have been shown very capable of many NLP-related tasks such as processing large body of text at once~\cite{llm-nlp-survey}, understanding the semantic meaning behind text~\cite{semantic-learners}, producing outputs in both user-readable format as well as machine-readable format~\cite{machine-readable}. 
Recognizing their capability, recent work  on privacy policy and compliance analysis have adopted LLMs in their workflow to perform some of these tasks rather than relying on previous NLP-based techniques~\cite{llm1,llm2}. 
However, it remains unclear whether LLMs can truly replace the diverse functionalities, methodologies and the manual rigor that was the norm in prior work in this domain. 
Understanding this is critical, given that it informs not only the future role of existing privacy policy analysis tools, but also how future research efforts in this domain should be allocated. 


To address this research gap, we perform the first systematic evaluation of privacy policy analysis tools from prior work against two state-of-the-art LLMs in their default configuration, mainly guided by the following three research questions:

\begin{itemize}
	\item[](\resq{1}) To what extent can LLMs {\textit replicate the functionalities} performed by tools developed by prior work?
	\item[](\resq{2}) To what extent can LLMs {\textit replicate the intermediate methodologies} that prior work relied on to design and implement their tools? 
	\item[](\resq{3}) What are the trade-offs of using LLMs to perform privacy policy analysis vs using traditional tools? 
\end{itemize}

Our analysis encompasses six different tools, covering three privacy policy analysis functionalities (contradiction detection in policy text, analysis of regulatory compliance, aggregation of multiple policies), and two intermediate methodologies upon which the current tools are built, the task of manual annotation by experts and Semantic Role Labeling (SRL)~\cite{srl}. 
We compare the performance of these tools against two state-of-the-art LLMs, contemporary models from Google (\ie Gemini-2.5-pro) and OpenAI (\ie GPT-5.2). 

Our findings (\fnumber{1}$\rightarrow$\fnumber{19}) reveal several key insights. 
First, our comparative analysis using a shared dataset of ten privacy policies reveals that LLMs can largely replicate the end-user functionalities of existing tools through prompting alone, surpassing the performance of the tools in almost every category. 
Second, even for challenging intermediate task such as manual annotation of privacy policies, LLMs show reasonable performance, producing metrics that are comparable to the agreement rates among human annotators. 
Third, our analysis reveals several trade-offs between the use of LLMs and prior tools in performing the analysis of privacy policies, with both having their own structural advantages and disadvantages that are not easily transferable to each other. Hence, we argue for a hybrid approach that balances the performance gains and flexibility of LLMs with the scalability and low-cost nature of prior tools. 

Our contributions can be summarized as follows: 
\begin{itemize}
	\item To the best of our knowledge, our work presents the first systematic evaluation of the capability of LLMs in replicating the functionalities and development methodologies of privacy policy analysis tools developed prior to the LLM era. 
	\item Our analysis covers six representative tools spanning three functionalities and two intermediate tasks, providing a comprehensive comparative study of LLMs against specialized tools in this domain. 
	\item We demonstrate that modern LLMs can successfully replicate not only end-user functionalities but also key intermediate tasks that have traditionally required considerable domain expertise and manual human effort. 
	\item Across most of the evaluated tasks, off-the-shelf LLMs can match or exceed the performance of the corresponding specialized tool, with only prompting strategies and no additional domain-specific configuration or training.
\end{itemize}
\section{Study Overview}
\label{overview}
In this study, our primary objective is to establish a baseline understanding of the extent to which LLMs can replicate both the diverse functionalities and the underlying methodologies of tools developed by prior work. 
While the former allows us to assess the feasibility of using LLMs as a general-purpose privacy policy analysis tool, the latter analyzes whether we can leverage the capabilities of LLMs as core building blocks of future tools, thereby avoiding the domain-specific limitations and engineering overheads of prior approaches (\eg intermediate representations of policy texts, manual labeling). 

To that end, our comparative analysis of the tools and the LLMs selected for this study is driven by the following goals, each informed by the corresponding \resq{}s: 

\myparagraph{\dgnumber{1} - Replicating Tool Functionalities using LLMs (\resq{1})}
Our first goal is to determine whether LLMs can replicate the functionalities that were previously performed by specialized tools. 
To achieve this goal, we first deploy and run each selected tool against a dataset of curated privacy policies. 
We then design tool-specific prompts to instruct LLMs to perform the same functionality while preserving the output format of each tool where possible \ie to enable a direct comparison of results. 
Both sets of results are then either compared against a manually constructed ground truth dataset, where applicable,
or via manual validation of the results. 

\myparagraph{\dgnumber{2} - Replicating Intermediate Analysis Tasks with LLMs (\resq{2})}
The development of many prior tools rely on the design and implementation of many intermediate NLP tasks such as manual data labeling (\ie to train models)~\cite{manual-labeling}, semantic role labeling~\cite{srl}, and other information extraction and classification techniques. These intermediate tasks are often necessary to transform unstructured privacy policy texts into structured representations so that the tools can perform the desired analysis on the structured representations (\eg contradiction analysis~\cite{Andow:SECURITY2019}, policy summarization~\cite{poligraph}). 
Through this design goal, we analyze the ability of LLMs to perform such intermediate tasks directly from the privacy policy text, thereby enabling us to develop tools without explicitly having to engineer and implement such tasks which were previously not only labor-intensive but also required significant domain expertise. 

To achieve this goal, we employ a similar analysis pipeline as \dgnumber{1} - we first identify tools that provide us these intermediate task results as outputs, and we design tool-specific prompts to instruct LLMs to generate the same output. 
By comparing the outputs of the tools with the LLMs, we assess how successful LLMs are in replicating these intermediate analysis tasks. 

\myparagraph{\dgnumber{3} - Trade-offs of LLMs vs Existing Tools (\resq{3})}
Even if LLMs can successfully replicate the functionalities of existing tools, which would mark a paradigm shift in this domain, switching to LLMs introduces new costs and benefits, signaling newer challenges and opportunities. 
Our third goal is to characterize the practical trade-offs associated with replacing existing, specialized tools with newer tools/techniques that rely on LLM-based workflows. To achieve this goal, we compare these two approaches along various dimensions such as deployment effort and cost, domain expertise requirements, scalability of the approaches, and output quality and ease of validation. 

%
%
%
%
%

\section{Methodology}
\label{sec:methodology-section}

\subsection{Dataset Construction}
\label{methodology-dataset}

\myparagraph{Tools Selection} 
Recall that our focus is to compare LLMs against {\em traditional} tools \ie tools that were developed prior to the advent of LLMs. Hence, we curated a list of tools that perform privacy policy or data compliance related analysis from the literature. 
For a comprehensive tool representation, we collected papers using a keyword-based search from major security and privacy conferences (\eg USENIX, IEEE S\&P, ACM CCS) between the years 2018 and 2023 \ie after GDPR was introduced and up to a year after ChatGPT was released to account for papers that may have been published during the pre- and post-LLM transition. This resulted in a set of 97 papers, out of which 27 proposed tools. 

To shortlist the tools that were applicable to our study, we established two deployment criteria that would make our analysis in this study feasible: 
i) the tools have to be deployable in our local machine (Macbook Air, M4 Chip, 24GB RAM), and 
ii) the tools have to take in our curated privacy policy dataset as input and produce output that is accessible to us locally. 
Applying these criteria to the full set of tools yielded a total of \textit{six tools}, which we further split across three end-to-end functionalities and two intermediate tasks.

Four tools could be analyzed across three end-to-end functionalities:
\begin{itemize}
	\item[i)] Intra-policy Contradiction Detection -- PolicyLint~\cite{Andow:SECURITY2019} and Poligraph~\cite{poligraph}
	\item[ii)] Regulatory Compliance Analysis -- AutoCompliance~\cite{autocompliance}, PolicyChecker~\cite{policychecker}, Poligraph~\cite{poligraph}
	\item[iii)] Policy Aggregation -- Poligraph~\cite{poligraph}
\end{itemize}
Note that Poligraph performs multiple functionalities and hence, it is included under each relevant functionality category in our study.  

The remaining two tools, PolicyPulse~\cite{policypulse} and Purpliance~\cite{purpliance}, enabled us to analyze a foundational intermediate task that serves as a building block of many prior privacy analysis tools: Semantic Role Labeling (SRL). Both these tools produce outputs for SRL that we can review, enabling such an analysis.


Finally, we also evaluate whether LLMs can perform another foundational intermediate task in prior work: Manual Annotation of policy text. For this evaluation, we leverage the OPP-115 dataset, a widely used corpus of expert-annotated privacy policies, as a benchmark against the annotation output produced by the LLMs. 

To summarize, our evaluation dataset comprises {\em six tools} and the OPP-115 annotated policies, covering {\em three end-to-end functionalities}: and {\em two intermediate tasks}.





\myparagraph{Privacy Policy Dataset Construction}
To enable a direct comparison of results, we curated a list of \textbf{\em ten privacy policies} associated with the top free apps (as of January 2026) from the Google Play Store, spanning six different categories: i) Productivity ({\em ChatGPT, Yingrun PDF Launcher}), ii) Social ({\em Tiktok, Instagram}), iii) Communication ({\em WhatsApp}), iv) Shopping ({\em Temu, Whatnot, Shein}), v) Video Players \& Editors ({\em Capcut}), and vi) Finance ({\em Cash App}). 
For each app, we extracted the corresponding privacy policy from the {\em ``Data safety''} section of their respective Google Play store page. 
Further, we convert the privacy policies into plain text using HtmlToPlaintext tool~\cite{Andow:SECURITY2019} to facilitate analysis by the tools and the LLMs. Note that given the length of the privacy policy text (\ie our ten selected privacy policies consisted of 1956 sentences and 52601 words) and the cost associated with processing large body of text with LLMs, we deliberately chose ten as a reasonable dataset size while providing representation across multiple categories to perform a generalizable, yet feasible comparison analysis, as we discuss in Section~\ref{case-for-tools}.

\myparagraph{LLM Selection} We selected GPT-5.2 and Gemini-2.5-Pro as the two LLMs for our evaluation, as they represented the state-of-the-art publicly available models at the time of this study (\ie January 2026), created and hosted by OpenAI and Google respectively. 
Both models are accessible via API, enabling automated evaluation across all ten privacy policies. 
Note that for simplicity, we refer to GPT-5.2 as simply \gpt, and Gemini-2.5-Pro as simply \gemini for the remainder of the paper.



\subsection{Analysis Pipeline}
\label{methodology-pipeline}
We analyze all ten privacy policies against the locally-deployed tools as well as the two LLMs, \gpt and \gemini.

\myparagraph{Analysis with Tools} 
We deploy each tool on our local workstation, provide the ten privacy policies in our test datasets as input, and collect the generated output. We then manually validate the result output to enable a comparison with the corresponding results produced by the LLMs. 

\myparagraph{Analysis with LLMs}
In line with our goal to assess whether LLMs can replicate the functionality provided by the tools, we design tool-specific prompts that provide LLMs with the instructions to perform the required analytical task on the same set of ten privacy policies provided as input. 
To enable a direct comparison with the tool output, the prompts also instructed the LLMs to provide the produced output in the same format as the corresponding tool's output where possible.

We employ two prompting strategies, {\em both single prompts}, to assess the overall impact of the provided prompts on LLM performance in these tasks: 
\begin{itemize}
	\item[i)] {\bf Simple prompt (\promptnum{1})} - A minimal prompt that asks the LLM to perform the corresponding functionality (\ie corresponding to the tool under comparison) without providing any additional context, definition or explanation. This prompting strategy represents a user with a limited domain expertise who simply wants the LLMs to perform the desired functionality and relies entirely on the LLM's pre-trained knowledge to interpret and perform the task. 
	\item [ii)] {\bf Detailed prompt (\promptnum{2})} - A comprehensive prompt that provides the relevant task description, domain-specific definitions, and any other contextual information to help facilitate the LLMs in performing the corresponding functionality. This prompting strategy represents a knowledgeable user who understands the functionality and the tool being replicated.
\end{itemize}

\myparagraph{Prompt Creation} 
For each tool, we designed prompts by first studying the tool's methodology, output format, and any domain-specific definitions or rules the output relied upon using the descriptions provided in the paper. Where possible, we purposefully include the {\em exact} functionality description, terminology definitions and usage that we extract from the paper into each of the prompts. 

In particular, for \promptnum{1}, we provide only the task description of the functionality that needs to be executed along with the expected output format (\eg the JSON key-values to be produced such that the output matches the format of the corresponding tool). As stated earlier, this strategy serves as a baseline non-expert user. 
For \promptnum{2}, in order to mimic a knowledgeable user who understands the functionality well, we include the functionality description, domain-specific context and any additional rules or definitions (\eg GDPR article requirements for AutoCompliance and PolicyChecker, CCPA term definitions for Poligraph). This strategy  is intended to mimic an expert user. 
We provide the details of the tool-specific prompt contextualization in Section~\ref{sec:func-replication}. An example prompt is included in the Appendix~\ref{app:detailed_prompts}, while the full policy prompt for all our experiments are provided in the online artifact~\cite{online-artifacts} due to space constraints. 

Additionally, where applicable, we instruct LLMs to provide the reasoning for each of their analysis decisions.
This is independent of any manual validation of results that we perform \ie we assess each produced result on its own merits without consulting the LLM's reasoning. 
Rather, this reasoning provides us with a supplementary opportunity 
to assess whether LLM's justification accurately supports its produced output, if needed. 


\myparagraph{Configuration of LLMs}
To minimize the variability of output and improve reproducibility,
all LLM experiments were conducted with the \texttt{\small temperature} parameter set to \texttt{\small 0} (~\cite{song2025good},~\cite{shi2024thorough}). 
As our goal in this study is to establish a baseline performance of state-of-the-art LLMs in this domain when compared with existing, specialized tools, we do not change any configurations (\eg we leave the `thinking' mode enabled) to allow for an off-the-shelf comparison. 



\myparagraph{Output Comparison} The outputs from both the tools and the LLMs were evaluated by comparing against the ground truth (where possible) or by performing manual validation. 

To elaborate, we construct ground truths for tasks where an exhaustive, deterministic set of correct answers could be derived from the privacy policy corpus (e.g., privacy policy aggregation in Poligraph~\cite{poligraph} which entails counting occurrences of specific data types across all privacy policies). The output from the tools and the LLMs is then compared against this ground truth to assess their quality. 
For open-ended tasks where this is not possible such as for contradiction detection (\eg PolicyLint~\cite{Andow:SECURITY2019}), two authors manually validate the outputs produced by both the tools and the LLMs, with all disagreements resolved through discussion. 
Since our goal is to establish a conservative baseline of LLM performance on these functionalities, any result 
that was deemed borderline after discussion was counted as a false positive. 




\section{Functionality Replication with LLMs (\dgnumber{1})}
\label{sec:func-replication}

We selected three functionalities (labeled \funcnumbers) for replication with LLMs-- Intra-policy Contradiction Detection, Regulatory Compliance Analysis, and Policy Aggregation --  based on tool availability within our deployment criteria (Section~\ref{methodology-dataset}), and compared the performance against the corresponding tools.
As we mention in Section~\ref{methodology-pipeline}, our goal here is to assess whether off-the-shelf LLMs, without any domain-specific fine-tuning or any additional engineering, can perform these functionalities at all, and to what degree their out-of-the-box performance holds up against specialized tools built explicitly for these purposes. 
However, to facilitate a direct comparison of outputs with the tools, we instruct the LLMs via prompt to produce results in the same format as the corresponding tool.

\subsection{Intra-policy Contradiction Detection (\funcnumber{1})}
\label{contradiction-detection}
This is one of the most common tasks in privacy policy analysis(~\cite{Andow:SECURITY2019, contradiction-pp}). 
The main objective of a tool performing this functionality is to detect logical contradictions and inconsistencies within the privacy policy text (\eg a privacy policy that states ``we do not collect any personal information'' and ``we collect your email for correspondence'' has a contradiction). 

\myparagraph{Experimental setup} As we specified in Section~\ref{methodology-dataset}, two tools in our dataset, PolicyLint and Poligraph, perform this functionality. Recall that we provide each tool with the ten privacy policies in our dataset as input to perform their analysis. However, Poligraph's contradiction analysis operates directly on PolicyLint's output rather than independently detecting contradictions, and it instead reclassifies PolicyLint's output to reduce false positives by incorporating additional context from the privacy policy (\eg differentiating between ``sell'' and ``share'' actions). 
Hence, per their documentation~\cite{poligraph}, we provide PolicyLint's contradiction results in addition to the privacy policies as input to Poligraph. As a result, we note that Poligraph's output is strictly bound by the detections made by PolicyLint and hence, no separate LLM replication was necessary to do a separate comparison with Poligraph \ie the same LLM outputs produced for PolicyLint serve as the upper-bound comparison for both tools.

For the LLM replication, we apply all two prompting strategies to both the LLMs. The prompts described the functionality \ie  identifying logically conflicting statement pairs, in increasing level of specificity, as we discuss in Section~\ref{methodology-pipeline}. The
In particular, for \promptnum{2}, we provide a precise definition of what qualifies as a contradiction as defined by PolicyLint.
The complete prompt can be found in the Appendix\ref{app:detailed_prompts}.

\subsubsection{Results} 
Given the open-ended nature of contradiction detection, we could not derive ground truth dataset from our privacy policy corpus. Instead, two authors manually validate the produced results from both the tools and the LLMs, achieving an overall agreement rate of 90.28\%.


\myparagraph{Tools vs LLM: PolicyLint and Poligraph} 
PolicyLint identified a total of five contradictions across all ten policies. Notably, all of them originated from a single policy (Shein), and it produced no output for the remaining nine policies. In our manual validation, we found all five contradictions to be false positives, {\em yielding an effective contradiction count of zero}. 
Recall that Poligraph reclassifies PolicyLint's output to further {\em narrow} contradictions by removing false positives. In a confirmation of our results, Poligraph returned zero contradictions when we provided PolicyLint's results as input. In summary, both the tools failed to produce any true contradictions across all ten privacy policies in our dataset. 

Unsurprisingly, following the poor benchmark set by the tools, all LLM-prompt configurations outperformed the tools. 
All configurations detected contradictions, with an average of 31.25 contradictions per configuration. \gpt produced the highest and the lowest number of contradictions under \promptnum{1} (\ie 44) and \promptnum{2} (\ie 19) respectively. 
To truly outperform the tools though, the true positive rate (\ie precision) matters. 
Upon manual validation, we found 18 true positive contradictions in total (out of 120 detected) across all configurations. \gpt-\promptnum{2} with a precision of 26.3\% was the best-performing, while \gpt-\promptnum{1} is the worst-performing with 9.1\% precision. Note that all configurations, {\em even \promptnum{1}},  produced valid contradictions, thereby outperforming both the specialized tools. 
We report the precision of all models and prompting strategies in Table~\ref{tab:perf_overview}.

\finding{1}{Both LLMs across all prompting strategies identified valid contradictions that the specialized tools entirely failed to detect, with the best-performing LLM configuration (\gpt-\promptnum{2}) achieving a precision of 26.3\% compared to the tools' effective precision of 0\%}

Of particular note, LLMs seemed to identify contradictions in highly sophisticated cases where traditional, \ie rule- or pattern-based tools, might struggle. 
See this contradiction identified by Gemini-2.5-Pro in Tiktok's privacy policy with \promptnum{1} and \promptnum{2} - {\em statement-1}: {\em "we do not sell... or share your personal information with third parties for purposes of cross-context behavioral advertising..."} and {\em statement-2}: {\em "we share information we collect with service providers and business partners as necessary... for business purposes, including.. advertising, marketing services..."}. Given that {\em statement-1} forbids the sharing of personal data for behavioral advertising, while {\em statement-2} shares information to business partners for purposes {\em including advertising}, which might subsume behavioral advertising as well, we mark this as a true positive. However, the relevant clause is hidden in {\em statement-2}, which was a single, long line with multiple embedded clauses, making it difficult for traditional tools to identify. 



Among the contradictions we validated to be true positives, none were detected across all models and prompting strategies. Out of the 18 unique true positive contradictions detected cumulatively by the LLMs, {\em 15 were unique to individual configuration} while only 3 were detected by at least two configurations. 
This suggests that no single model and prompting strategy exhaustively captured all {\em true} contradictions, 
and hence,
using different prompting strategies (\ie providing different context in prompts) can yield more comprehensive coverage during contradiction detection than any individual strategy alone. 

\finding{2} {Unlike traditional tools, LLMs may produce variable output and distinct contradictions across different prompting strategies. Hence, combining different LLM and prompting strategy can result in more comprehensive coverage of contradictions}

Further, while \promptnum{1} consistently produced higher number of contradictions in both models, it also achieved {\em significantly lower} precisions when compared with \promptnum{2}, 17.1\% vs 25.0\% for \gemini and 9.1\% vs 26.3\% for \gpt. 
This indicates that detecting true contradictions requires the model to apply precise, domain-specific criteria for what constitutes a genuine contradiction, included in \promptnum{2}. 
Without this added context, the model resorts to flagging any superficial differences between sentences as contradictions. For example, one of the contradictions flagged by \promptnum{1} but {correctly} not flagged by \promptnum{2} was found in WhatsApp privacy policy. In this sentence pair, the first sentence listed {\em ``WhatsApp Ireland LTD''} as the service provider for EU users, while the second sentence directed users to issue any privacy inquiries to {\em ``WhatApp LLC''} in the US, which is not a contradiction. Since \promptnum{1} lacks any definition of what qualifies as a contradiction, it potentially flagged the different entity names as a conflict. 

\finding{3} {Without precise contradiction criteria progvided as context, LLMs can flag superficial differences between policy statements as a contradiction, significantly increasing false positives}
%


%

\subsection{Regulatory Compliance Analysis (\funcnumber{2})}
\label{compliance-analysis}
A tool's main objective in executing this functionality is to analyze whether an input privacy policy satisfies the disclosure requirements mandated by a given regulation that the tool supports (\eg CCPA~\cite{ccpa}, GDPR~\cite{gdpr-special}). As we specified in Section~\ref{methodology-dataset}, three tools in our dataset perform this functionality: AutoCompliance and PolicyChecker support GDPR while Poligraph supports CCPA. 
However, given the relatively large scope of both CCPA and GDPR, each of these tools target a specific regulatory scope within the regulation they support. 
In particular, AutoCompliance and PolicyChecker both analyze GDPR completeness but AutoCompliance checks for compliance against GDPR Article 13 (\ie a requirement that requires companies to disclose to users how their data is collected and processed at the time of collection~\cite{gdpr13}), while PolicyChecker verifies completeness against GDPR Articles 13 and 14 (\ie additionally covering cases where personal data is indirectly collected from third parties rather than from the user directly~\cite{gdpr14}). 
Alternately, Poligraph analyzes privacy policies to identify any inconsistencies in how the policy text defines or categorizes specific data objects in relation to how CCPA defines or categorizes them. For instance, a privacy policy classifying an IP address as non-personal information would be flagged as a violation since CCPA defines it as personal data~\cite{ccpa}. 

Although AutoCompliance and PolicyChecker both analyze privacy policies for GDPR violations, they employ different underlying techniques: AutoCompliance deploys a supervised learning-based classification model to classify policy sentences~\cite{autocompliance} while PolicyChecker relies on a rule-based, semantic role labeling approach~\cite{policychecker}. Further, while AutoCompliance checks for the presence of GDPR Article 13 provisions, PolicyChecker performs a much granular analysis for Article 13 and 14 provisions, distinguishing between mandatory requirement violations (\ie disclosures that {\em must} be present the policy text without exception \eg data controller's identity, rights of data subjects), legal basis violations (\ie policy text failing to provide a legal justification for processing personal data \eg user consent), and conditional requirements (\ie disclosures that become mandatory based on context \eg identity of entities receiving data if data is shared). 
On the other hand, Poligraph employs a graph-based approach to extract data objects from privacy policies into local ontologies and flags inconsistencies by comparing them against a global CCPA-based data ontology. Like PolicyChecker, it performs a more granular, multi-dimensional analysis on the input policy text \ie it identifies both i){\em non-standard terms} which are terms used within the policy text but not defined in the CCPA global ontology (\eg vague terms like ``technical information''), and ii) {\em misleading definitions} which are terms that are defined in the privacy policy under a category different to the CCPA ontology (\eg classifying IP-address as non-personal information whereas CCPA considers this personal information). 

\myparagraph{Experimental Setup} 
While our goal (\ie under \dgnumber{1}) is to compare the tool performance against LLMs at the {\em functionality} level rather than the methodology level, we provide additional, methodology-related contexts in \promptnum{2} if they are necessary for LLM to perform the corresponding functionality \eg we provide the global CCPA-based data ontology in \promptnum{2} when replicating Poligraph, we provide the definitions of mandatory requirement, legal basis and conditional requirements in \promptnum{2} when replicating PolicyChecker. The full prompts, both \promptnum{1} and \promptnum{2}, for all three tools are available in our online artifact~\cite{online-artifacts}. 
As before, the tools themselves are treated as black boxes, \ie we execute the same ten privacy policies as input in all three tools and evaluate their outputs without any regard to their internal differences. 


\subsubsection{Results}
For comparison with AutoCompliance and PolicyChecker, we manually create ground truth datasets by analyzing the ten privacy policies along the GDPR Article 13 and 14 provisions that are under the scope of the tools. We compare the results of the tools and the LLMs against this ground truth dataset. 
For Poligraph, as ground truth dataset construction was infeasible, we manually validate the output produced by the tool as well as by the LLMs exhaustively, achieving an overall agreement of 89.07\%.
 

\myparagraph{Tool vs LLM: AutoCompliance}
All LLM-prompt combination outperformed AutoCompliance, as seen in Table~\ref{tab:perf_overview}. In comparison with the ground truth, AutoCompliance achieved an accuracy of 78\% but an overall F1 score of only 48\% due to both precision (59\%) and recall (40\%) being low. 
In contrast, across all metrics of accuracy, precision and score, each of the LLM-prompt combinations performed better. Surprisingly, \gpt-\promptnum{1} achieved the highest overall score (F1-94\%) while \gpt-\promptnum{2} (F1-92\%) and \gemini-\promptnum{2} (F1-90\%) are close behind. The worst-performing \gemini-\promptnum{1} (F1-85\%) still substantially outperforms the tool (F1-48\%). 

\finding{4}{LLMs significantly outperform AutoCompliance in the regulatory compliance analysis, with the best performing configuration achieving an F1-score of 94\% compared to the tool's F1-score of 48\%.}

The tool's reliance on patterns learned from its training corpus may cause it to miss requirements expressed using non-standard terms or phrasing, whereas LLM was able to identify even implicitly stated requirements. For example, for a policy that stated {\em ``we take steps to verify your identity before granting you access to your personal information''}, AutoCompliance failed to identify the implicit granting of access to data, while LLM correctly identified such cases. 

\finding{5}{LLMs can recognize GDPR requirements expressed implicitly or through non-standard phrasing, which overcomes a key limitation pattern-based classification and demonstrating a semantic understanding of the policy text rather than simply relying on lexical matches.}.

\myparagraph{Tool vs LLM: PolicyChecker} 
All LLM-prompt combinations except \gpt-\promptnum{1} outperformed PolicyChecker, as seen in Table~\ref{tab:perf_overview}. 
In comparison with the ground truth, PolicyChecker achieved an overall F1-score of 68.7\%. While the recall (83.9\%) is reasonable (\ie when compared to the results of the LLMs), the precision (73.3\%) and accuracy (37.7\%) are quite low, especially when compared to \promptnum{2} for both LLMs. 
In contrast, the best performing LLM-prompt combination, \ie \gemini-\promptnum{2} achieved an F1-score of 89.9\%, with a high precision (92.9\%) and recall (92.9\%) and a reasonable accuracy (73.1\%).

\finding{6}{LLMs significantly outperform PolicyChecker in the regulatory compliance analysis, with the best performing configuration achieving an F1-score of 89.9\% compared to the tool's F1-score of 68.7\%.}

Of particular note, both models perform significantly worse in \promptnum{1} strategy than \promptnum{2}. This gap in PolicyChecker is more pronounced than in AutoCompliance, potentially given the increased complexity and scope of PolicyChecker's analysis. Recall that PolicyChecker not only covers more regulatory provisions (\ie both Article 13 and 14 of GDPR), it also performs more granular analysis compared to AutoCompliance (\ie including mandatory requirement and legal basis violations). 
Without the contextual information provided in \promptnum{2}, LLMs lack the precise criteria to correctly assess a violation, resulting in lower accuracy and precision, as seen in Table~\ref{tab:perf_overview}. 

\finding{7}{\promptnum{2} significantly outperforms \promptnum{1} strategy in both LLMs, suggesting that when tasks are more granular and context-dependent, the more critical it is to provide LLMs with domain-specific criteria in the prompt.}


\myparagraph{Tool vs LLM: Poligraph} 
Recall that, for any input privacy policy, Poligraph identifies both the {\em non-standard terms} and the {\em misleading definitions} per the CCPA's categorization.  
For the ten privacy policies in our dataset, Poligraph identified six potential inconsistencies, all six for {\em misleading definitions} and zero for {\em non-standard terms}. 
Upon manual validation, four of them were confirmed as true positives, resulting in a precision score of 66.7\%.

All LLMs produced significantly higher number of potential inconsistencies,  with \gpt with both \promptnum{1} and \promptnum{2} producing a significantly higher total than the rest. 
Our manual validation of the results showed that \gemini-\promptnum{1} achieved the highest precision of 75\% (\ie 48/64 true positives), outperforming Poligraph's precision of 66.7\% even after producing a significantly larger number of inconsistencies (see Table~\ref{tab:perf_overview}).

\finding{8}{\gemini outperformed Poligraph, producing a higher number of potential inconsistencies, higher number of true positives and a higher precision score}

While \gpt achieved lower precision, it produced the highest number of true positives, indicating substantially broader coverage of inconsistencies than either Poligraph or \gemini. 
However, the relationship between precision and coverage is not uniform across LLMs. \gemini achieved higher precision than Poligraph while still identifying more true positives, albeit fewer than \gpt.

This result highlights a fundamental advantage of LLMs over traditional tools. Poligraph adopts a conservative, pattern-based, ontology-driven methodology that might reduce false positives but it comes at the cost of missing inconsistencies that fall outside its predefined rules. 
By contrast, LLMs can perform semantic and contextual reasoning within policy text, enabling them to recognize data objects and their inconsistencies beyond predefined rule-sets (\eg CCPA ontology). Hence, our results indicate that LLM-based approaches can achieve substantially greater coverage and identify higher number of {\em valid} inconsistencies, while still maintaining a comparable, or in some cases, improved precision score. 
For example, consider this from the Capcut privacy policy. {\em ``..we may receive information about you from payment service providers, such as payment confirmation and purchase details''}. The phrase {\em payment service providers} was correctly flagged as a {\em non-standard term} by \gpt but was missed by Poligraph. Similarly, LLMs correctly identified non-standard terms such as ``transaction history'', ``survey data'' that were missed by Poligraph. 


\finding{9}{The ability to interpret privacy policies semantically rather than through predefined patterns enables LLMs to identify substantially higher inconsistencies, yielding substantial gains in coverage and true positives without a corresponding compromise in precision.}


\subsection{Policy Aggregation (\funcnumber{3})}
\label{policy-summarization}

Unlike \funcnumber{1} and \funcnumber{2}, both of which analyze privacy policies individually, a tool executing \funcnumber{3} extracts and consolidates information across multiple privacy policies into a structured, aggregated representation, facilitating a multi-policy analysis of how data objects are collected, processed and shared, and enabling insights into common data practices across the industry.  
As discussed in Section~4.1, Poligraph is the only tool in our dataset that provides this functionality. 

\myparagraph{Experimental Setup}
As before, we used Poligraph as a black box and provided it with the ten privacy policies in our dataset. We then ran the ``Policies Summarization'' function within the tool (\ie since it supports multiple functionalities)~\cite{poligraph}. 
The tool then produces aggregated numbers regarding data practices across all privacy policies, such as the types of data collected, entities collecting the data and purpose of collection. 
Similarly, for LLMs, we create \promptnum{1} by specifying the functionality (\ie to extract data collection information, entities and purposes across all privacy policies), whereas in \promptnum{2}, we provide additional details such as the data types, entity categories and purpose categories as defined by the Poligraph tool. The full prompts are provided in our online artifact~\cite{online-artifacts}.

\subsection{Results}
To enable comparison, we manually create a ground truth dataset by analyzing each privacy policy and extracting the frequency and source sentence of all the data types, entities and purpose classes exactly as defined by Poligraph \eg counts for data types such as contact information, government identifier, geolocation, entities such as the first party, advertiser, content provider and purposes such as security, legal or advertising. 

\myparagraph{Tool vs LLM: Poligraph}
In line with our observations before, all LLM-prompt combinations significantly outperformed Poligraph in this functionality. Poligraph achieved an F1-score of 67.1\% while the best performing LLM, \ie \gpt-\promptnum{2}, achieved an F1-score of 88.2\% and even the worst performing LLM, \ie \gemini-\promptnum{2}, performed significantly better with an F1-score of 81\%.

\finding{8}{All LLM configurations outperformed Poligraph in the policy summarization and aggregation task, with the best performing LLM achieving an F1-score of 88.2\% compared to Poligraph's 67.1\%.}

Interestingly, unlike any other tool across all other functionalities, Poligraph was able to outperform the LLMs in one particular metric. It achieved a precision of 98.2\%, outscoring all LLMs (see Table ~\ref{tab:perf_overview}), although its poor recall and overall accuracy resulted in a low F1-score. 
This performance is notable given that Poligraph uses predefined patterns and ontology-based mappings to perform this task, which allows it to identify objects with a high confidence but at the expense of flexibility in recognizing implicit or lexically different yet semantically similar patterns, leading to lower recall. 
In contrast, the best LLM-prompt combination still produced a comparable precision of 93.4\% while maintaining a substantially higher recall, indicating the robustness of LLMs both in handling different text variations while producing a good precision without relying on predefined rules. 

\finding{9}{While traditional techniques such as pattern matching can lead to high precision, it might come at the sacrifice of achieving a good recall. LLMs can balance this trade-off by maintaining a comparable precision while still accounting for text variations that go beyond the scope of pattern recognition}

\section{Intermediate Task Replication with LLMs (\dgnumber{2})}
\label{sec:method-replication}
While \dgnumber{1} evaluated the ability of LLMs to replicate end-user functionalities of tools, these tools are reliant on intermediate NLP tasks to process unstructured policy text and transform it into structured representations to execute the functionality-level analysis. Such tasks often require manual labor or domain expertise and can manifest as onerous engineering tasks during development. 
Hence, automating them can be hugely beneficial to reduce burden on developers, and to scale the ability of the tools, thereby enabling researchers to build better tools on top of these intermediate tasks. 

In this analysis, we thus attempt to understand whether off-the-shelf LLMs can perform these intermediate tasks directly. To evaluate this capability, we select two representative intermediate tasks that are widely used in privacy policy analysis literature, and upon which many of the tools in this analysis are dependent on: i) Manual Annotation of policy text, ii) Semantic Role Labeling (SRL).

\subsection{Manual Annotation}
\label{sec:manual-annotation}
Many prior tools rely on manually annotated datasets to train and evaluate their own models\cite{hfl18, opp115-use1}. 
Creating such datasets has often been a very costly, labor-intensive task requiring significant domain expertise to interpret policy text and assign labels based on predefined criteria\cite{opp115, manual-dataset}. 
Hence, this task has frequently been the bottleneck during the design and development of tools in prior work. 

Unlike previous tasks, where we could compare the outputs of LLMs directly with the outputs produced by a tool, this analysis requires us to have a human-annotated dataset as a reference against which the LLM-generated annotations can be evaluated. 
Hence, we perform this analysis using the OPP-115 dataset~\cite{opp115}, a popular dataset that was manually-labeled by domain experts and has been extensively used in privacy policy analysis research~\cite{opp115-use1, opp115-use2}. 

\myparagraph{Experimental Setup}
For a like-for-like comparison, we selected ten policies from the OPP-115 dataset based on their global popularity ranking provided by Usable Privacy~\cite{usable-privacy}. 
To keep the analysis tractable across all selected policies, we focus on two annotation categories- {\em Category-A}: First-Party Collection and {\em Category-B}: Third-Party Sharing, which represent two of the most fundamental disclosure requirements in GDPR, \ie under Article 13 and 14~\cite{gdpr13, gdpr14}.

For each category, we create two prompts, \ie \promptnum{1} and \promptnum{2} with varying level of details and context to reflect the requirements of the labeling task, similar to our prompting strategy in Section~\ref{methodology-pipeline}. 
To be precise, \promptnum{1} instructs the LLM to label all segments as applicable within the privacy policy to the category under analysis: either {\em Category-A} or {\em Category-B}, as well as all the associated attributes and their values within that category. 
For example, for {\em Category-A}, \promptnum{1} instructs the LLM to label segments into the allowed category `First-Party Collection', and produce other associated attributes such as `Collection mode' (\ie implicit/explicit/unspecified), and `Personal Information Type' (\ie Financial/Health/Contact/Location). In \promptnum{2}, we augment the details in \promptnum{1} to add definitions for the category as well as for all attributes. The complete prompts are provided in our online artifacts~\cite{online-artifacts}. 

\subsection{Results}
Both LLMs are provided with the ten privacy policies as input, along with instructions in \promptnum{1} and \promptnum{2} for each category, resulting in a total of 80 queries (\ie 2 LLMs * 2 prompts * 2 categories * 10 privacy policies). 
We compare the annotations generated by the LLMs for each category against the original OPP-115 labels and compute the precision, accuracy, recall and the combined F1 score. 
Note that we compare our results against the `consolidated-1.0'  dataset provided within OPP-115\cite{opp115}, which represents the combined results produced by ten labelers in OPP-115.

\myparagraph{LLM vs OPP-115}
\gemini-\promptnum{1} achieves the strongest agreement with human annotators in OPP-115 among all configurations, with an overall F1 score of 79.4\%. 
Between the two models, \gemini achieved a higher F1 compared to both \gpt configurations (\ie 79.4 vs 70.2 for \promptnum{1} and 77.5\% vs 74.0\% for \promptnum{2}). 
Of note is that both models improved their recall when the prompts were supplemented with definitions of the categories, indicating that adding explicit category and attribute definitions in the prompt helps LLMs to identify a larger fraction of the labels present in the OPP-115 ground truth. However, this came at some precision trade-off for \gemini, and adding definitions (\ie \promptnum{2}) is not a guarantee to improve the overall performance, as our results show in Table~\ref{tab:perf_opp115_thinking}. 

Per category, LLMs perform much better for {\em Category-B} (\ie third-party sharing), with the best performing configuration \gemini-\promptnum{1} achieving 84.5\% F1 score compared to the best performing configuration for first-party collection (\ie \gemini-\promptnum{2}) which has an F1 score of 72.7 (full results in Table~\ref{tab:opp115_category_metrics} in Appendix).
One potential reason for this discrepancy could be that third-party sharing sentences are often stated in a standalone section, explicit and non-ambiguous manner (\eg ``We may share your information with..'') while first-party collection sentences can be implicit and scattered throughout the policy text (\eg ``if you contact our support team, we may request additional details to help resolve your issue, such as your device information or order history''). 

\finding{10}{The best-performing LLM for {\em third-party sharing} significantly outperforms the best-performing LLM for {\em first-party collection} by 11.8\%, indicating that first-party collection sentences are inherently harder to identify for LLMs than third-party sharing sentences.}

Note that during the creation of the OPP-115 dataset, the labelers achieved a cohen-kappa agreement score of 76\% for each of these categories~\cite{opp115}. This suggests that while the best F1 score for {\em third-party sharing} outperforms the human agreement score in OPP-115, the F1 score for {\em first-party collection} lags slightly behind, suggesting that LLMs may not yet replace expert human annotators for some of the challenging labeling categories. 

\finding{11}{LLMs can achieve an F1 score that is at an acceptable agreement rate among human annotators for {\em third-party sharing}, while for {\em first-party collection}, which may be more challenging to label, their F1 score lags inter-human labeling agreement in OPP-115.}


Further, 
we also compare LLM performance against the original human labelers from the OPP-115 dataset. 
OPP-115 was produced by ten labelers in total, with multiple labelers annotating multiple policies but no single labeler annotating all, as that would significantly increase the workload and potentially impact the quality of the dataset. 
The ten privacy policies in our analysis were labeled by nine (out of the ten) labelers, with individual workloads ranging from a minimum of one policy to a maximum of six. 

Table~\ref{tab:opp115_per_human} (in Appendix) shows the distribution and our results against each human labeler, using the same identifiers as OPP-115 (\eg \#82, \#84)
 to enable a direct comparison.
\gemini-\promptnum{1} consistently achieves an F1 score over 70\% for all labelers except \#121, with a maximum  80.3\% for the policy set analyzed by \#82. Consistent with our earlier observation, adding definitions (\ie \promptnum{2}) show mixed results, with performance improving slightly for \gpt while degrading a bit in \gemini. 

Notably, \gemini-\promptnum{1}'s F1 scores outperforms the agreement score against four of the nine labelers and stagys comparative for the rest, while \gpt outperforms two human labelers out of the nine. This comparison is interesting, given that the task of labeling itself is inherently  subjective and no two human labelers achieved a 100\% consensus even in the OPP-115 dataset, \ie the ground truth for this comparison. Hence, if a baseline level of noise between human annotators is acceptable, then the comparison of LLM annotation should also be  whether its disagreement with human labels falls within the acceptable range. By this standard, the results from \gemini are equivalent to adding a new human annotator to the task. 

\finding{12}{The F1 scores of \gemini-\promptnum{1} exceeds the agreement rate in the OPP-115 dataset for at least four human annotators, indicating that automated labeling using LLMs is comparable to adding an extra human annotator.}

\subsection{Semantic Role Labeling Task}
\label{sec:srl}
Semantic Role Labeling (SRL) is a foundational intermediate NLP task that  many privacy policy analysis tools rely on to transform unstructured policy text into a structured format so that a functionality-level analyses (\eg \funcnumber{1}$\rightarrow$\funcnumber{3}) can be performed on them. 
At its core, SRL decomposes a sentence into its underlying semantic structure, and identifies semantic relationships between objects in the sentences. 
For example, in the sentence ``We collect email address..'', SRL identifies ``We'' as the agent performing the action, ``collect'' as the action, and ``email address'' as the object on which the action is performed. 
Such a decomposition is a prerequisite to enable any functionality-level analysis  on the policy text at scale.
We select this intermediate task given that it evaluates whether LLMs are able to replicate a core NLP task that many of the traditional tools relied on for enabling their privacy policy analyses. 

\myparagraph{Experimental Setup}
Purpliance leverages 
SRL to decompose policy sentences into structured tuples in the form of (first/third party, action, data, purpose). 
Alternately, PolicyPulse leverages an SRL-based model 
to first identify relevant clauses in the sentences and map them into relevant data-practice categories such as first/third party and purpose as key-value pairs.  

For both tools, Purplicance and PolicyPulse, we applied the same two prompting strategies established before in Section~\ref{methodology-pipeline}, across the two LLMs, \gpt and \gemini. In \promptnum{2} for each tool, we provide the context and definitions of the data categories derived from the corresponding tool documentation. 
The full prompt is available in our online artifact~\cite{online-artifacts}. 

\subsection{Results}
We first ran both tools and all LLM-prompt combinations across all ten privacy policies in our dataset, producing the full set of tuples (\ie for Purpliance) and key-value pairs (\ie for PolicyPulse). 
Due to the large volume of generated output (\ie 1134 total sentences resulting in 6143 total tuples and key-value pairs to analyze), exhaustive manual validation over the full dataset was infeasible. 
Hence, we perform validation using two strategies: 
i) to ensure breadth, we perform a stratified sampling of 10\% of the results to get a representative sample, resulting in 118 sentences and 697 tuples selected for manual validation, and 
ii) to ensure depth, we select one privacy policy with a large-but-feasible output count, Tiktok with 142 sentences and 1127 total tuples and key-value pairs, to validate its results exhaustively. 
Further, we analyze the SRL performance across four important data classes in all of the outputs: the {\em data type} (\eg name), {\em action} performed (\eg collect), {\em receiver} of the data (\eg first-party), and the {\em purpose} (\eg account creation). 

Two authors performed the manual validation of the sampled outputs, resulting in 0 disagreements.

\myparagraph{Tool vs LLM: Purpliance}
In the manual validation of the 10\%  sampled output taken across all privacy policies, \gemini-\promptnum{2} performed significantly better than Purpliance, achieving a precision of 96.2\% over Purpliance's 79.8\%. 
Against all other LLM-prompt combinations, Purpliance's performance either matched (\ie against \gemini-\promptnum{1}, 79.9\%) or slightly exceeded the LLM performance (\ie against both \gpt prompts, 76.7\%). 


However, the gap becomes much starker when we manually validate the results over a full privacy policy (\ie Tiktok's). Here, all LLM-prompt combinations outperformed Purpliance, with the best performing model again, \gemini-\promptnum{2} bettering Purpliance's precision by 15.4\%. 

\finding{13}{LLMs either matched or vastly  outperformed Purpliance in the Semantic Role Labeling task, with the best performing \gemini-\promptnum{2} outscoring Purpliance's precision by more than 15\%}. 

Looking deeper into the performance on individual data classes, the performance of all LLMs is slightly  more uniform across all data  classes (see \figref{fig:purpliance_field_precision_heatmap} in Appendix \secref{app:srl_field_precision}), while 
Purpliance significantly struggles to identify {\em purpose} class with a precision of only 55\%, while achieving over 84\% in other classes. Although LLMs also struggle with the {\em purpose} class more, their performance variance is less severe. 

\finding{14}{Purpliance struggled to identify the {\em purpose} class much more than other data classes, while otherwise largely mirroring the performance of LLMs.}



\myparagraph{Tool vs LLM: PolicyPulse}
PolicyPulse on the other hand performed much poorer against the LLMs. 
Across the validated sample, its precision lagged nearly 10\% behind even the poorest performing LLM (\ie 65\% vs 55.4\% against \gpt-\promptnum{2}), while the best performing LLM-prompt combination, \gpt-\promptnum{1} outscored it by nearly 33\% with an overall precision of 88.1\%. 

The same trend holds even for the results over the Tiktok's privacy policy, with all LLM-prompt combinations significantly outperforming PolicyPulse. 
Here, the best performing \gpt-\promptnum{1} had a precision of 91.1, and outperformed PolicyPulse (59.9\%) by 31.2\%. 

\finding{15}{All LLM-prompt combinations comfortably outperform PolicyPulse, indicating LLM's strong compatibility in performing SRL task.}

Results on individual data classes identified by PolicyPulse and the LLMs again reveal that the performance of all LLMs is much more uniform across all data  classes (see \figref{fig:policypulse_field_precision_heatmap} in \secref{app:srl_field_precision}), while PolicyPulse varies widely, ranging from 44\% for the {\em receiver} to 67\% for the {\em purpose}. 

\finding{16}{While the performance of PolicyPulse across the identification of different data classes varied widely, LLMs remained much more consistent, indicating an ability to 
generalize regardless of class-specific complexities.}
%
%
%

\section{Trade-offs of LLMs vs Existing Tools (\dgnumber{3})}
\label{sec:cost-analysis}

While Sections~\ref{sec:func-replication} and ~\ref{sec:method-replication} demonstrated to what extent off-the-shelf LLMs can successfully replicate both the end-user functionalities as well as the intermediate tasks performed by traditional tools (\ie \resq{1} and \resq{2}), we discuss the trade-offs between both approaches in this section, given that they differ substantially in terms of deployment requirements and overall cost of use, extensibility and the robustness of the produced results. 
While answering \resq{1} and \resq{2} required quantitative performance metrics, we aim to provide key insights based on our experimental observations throughout this study to answer this \resq{3}. 




\subsection{Case {\em for} LLMs}
\label{case-for-llms}

Beyond raw performance metrics that we observed earlier, our  analysis surfaced several, {\em structural} advantages in LLM-based privacy policy analysis. 

\myparagraph{Lower Barrier of Entry} We deliberately designed \promptnum{1} to mimic a baseline user that provided only high-level description of the desired functionality to the LLMs, and did not provide step-wise instructions on how to execute the task. 
Across all functionalities, at least one of the LLMs outperformed the performance of the corresponding tool using only \promptnum{1} (see Table~\ref{tab:perf_overview}). This suggests that a non-expert user who understands the objective of the analysis but lacks any specialized knowledge of privacy policy or regulatory compliance analysis is able to exceed the performance of any specialized analysis tools simply by providing basic prompts to the LLMs. 

\finding{17}{LLMs receiving only a high-level description of a functionality can outperform specialized analysis tools, significantly lowering the barrier of entry for tool development that has traditionally required domain expertise and non-trivial engineering cost.}

\myparagraph{Improved Auditability of Results} 
At first glance, it might appear that the higher recall achieved by LLMs across all functionalities, while desirable, creates an additional burden given that these outputs may have to be validated. 
However, an important advantage of LLMs over traditional tools is that they can be trivially instructed to provide supporting evidence alongside the results they produce, a non-trivial task that requires design or implementation changes in traditional tools. For example, we instructed LLMs to provide the location of a sentence where an inconsistency was found, or provide reasons for a contradiction claim. This enables any output to be validated by tracing back to the underlying policy text.

While we did not use this as the primary basis for our manual validation, this advantage was particularly meaningful when performing validation in \funcnumber{3} \ie Policy Summarization and Aggregation. 
Recall that to enable comparison with LLMs in this task, we constructed the ground truth dataset by manually identifying data objects across all privacy policies, and recording both their frequencies and source sentences (see Section~\ref{policy-summarization}. 
While the results from both the tool and the LLMs did not match the ground truth exactly, the additional details in the LLM results allowed us to ascertain the cause of this divergence from the ground truth, which was infeasible to achieve with the tool. 

\finding{18}{Unlike traditional tools, which typically produce only their final outputs, LLMs can be instructed to provide provenance information depending on the functionality, making their results substantially easier to validate and inspect.}

\myparagraph{Generalizability Across Functionalities}
In our experiments, a single LLM when given functionality-specific instructions through prompts could perform all three functionalities at a level that either compared with or exceeded the performance of the specialized tools. This significantly reduces the effort required by policy experts and researchers to perform analyses along multiple functionality dimensions, while also reducing the engineering cost associated with developing and maintaining multiple functionality-specific frameworks. 
 Additionally, LLMs are also not bound by the common limitations that exist in traditional tools such as the pattern-based analysis, limited training done on a smaller corpus or the need to perform retraining to adapt to newer regulatory language or novel terminology (see \fnumber{5}). 

\finding{19}{LLMs not only adapt to multiple functionalities, but are also not bound by many limitations that exist in prior tools, making them substantially more agile to adapt to changing regulations and privacy landscape}

%

\subsection{Case {\em for} Tools}
\label{case-for-tools}

\myparagraph{Scalability of analysis}
While we incurred some deployment challenges for some of the tools (\eg workaround for specific hardware requirements), once deployed, the tools incurred no additional ``per-query'' cost, and was run entirely on local hardware, as specified in Section~\ref{methodology-pipeline}. 

This is a significant advantage over contemporary LLMs. Currently, state-of-the-art LLMs (\ie those used in this study) are not open-sourced, and are accessible only through paid APIs, resulting in a recurring cost that might increase based on the number and the size of the query sent in each request. 
Consequently, the cost of analysis increases rapidly with the dataset size and the analysis being performed, particularly if analysis involves multiple models or needs to be repeated. Local deployment of LLMs could help avoid API usage costs, but this is often prohibitively expensive, and requires high-cost, dedicated hardware that may be infeasible for most users to acquire. 

Recall that our analysis is done on a relatively small dataset  
of ten privacy policies, with cumulatively 1956 total sentences and 52601 total words. 
Nevertheless, this study incurred a cost of about \$350 in total, for all experiments run across both LLMs, including supplemental experiments that we ran to observe the effects of prompt sensitivity, that we discuss in Section~\ref{sec:limitation}. 
Using traditional tools, this cost can be entirely avoided, and the analysis can scale to hundreds or thousands of privacy policies with negligible additional cost.

\myparagraph{Stable Performance}
Our results show that LLM performance can vary substantially based on prompting strategy alone (\eg \fnumber{3}, \fnumber{7}). 
For each LLM, as seen in Table~\ref{tab:perf_overview}, we observed changes in performance across the replication of all functionalities and intermediate tasks under \promptnum{1} and \promptnum{2}. 
As discussed elsewhere in the literature~\cite{prompt-sensitivity}, such sensitivity in LLM performance also impacts the reproducibility of research (similar to ours). 
While the tools produced worse performance, they do not exhibit similar input-sensitivity, which makes their performance behavior easier to characterize. 

\myparagraph{Auditability of Failures} 
When traditional tools fail at producing a particular output, their results can be traced back to the specific component or ruleset within the tool that is responsible (\eg during our manual evaluation, we were able to associate the lower recall in many tools to their rigid, pattern-based design). This is generally possible because their design    constitutes discrete, modular components (\eg SRL Parsers, ontology mappings). 
By contrast, LLMs are opaque, monolithic systems where such kinds of debugging is infeasible to perform (\eg we are not able to map our observation in \fnumber{2} to specific internal components within the LLMs), making the process of developing tools over multiple iterations particularly challenging. 
Hence, in contexts where understanding {\em why} a failure occurred may be critical  to understand (\eg in regulatory contexts), traditional tools may still be preferable as they allow for greater control and auditability of their internal workings.  

\myparagraph{Advantages of High Precision} Multiple tools, with an exception of AutoCompliance and PolicyLint, 
achieved a comparable precision score compared with the LLMs. 
In one notable case (\funcnumber{3}), Poligraph outperformed the LLMs in precision, albeit with a much lower recall resulting in a far lower F1 score. 
This indicates that traditional tools can still serve some  important functions such as working as a high-precision filter to filter out false positives in a large output set, or as a component in a hybrid pipeline to constrain (or validate) the open-ended outputs from LLMs. 


\myparagraph{Developing Future Tools}
While our findings in this study, at first glance, might suggest that all future tools developed in this domain should leverage LLMs, our observations were mixed.  
One one hand, 
LLMs offer some critical advantages that removes some of the most important limitations from prior tools (\ie less reliance on domain expertise, general-purpose tool to encompass many functionalities, providing support in manual labeling), many important advantages of prior tools are difficult to materialize in LLMs currently. 
As we discussed in Section~\ref{case-for-tools}, deployed tools incur negligible (if any) per-analysis cost, can run entirely on local hardware, can scale the analysis to any size, and provide the ability to trace failures to design choices that are almost impossible to replicate in current LLMs. 
Hence, we argue that rather than using LLMs as a wholesale replacement to traditional systems, a hybrid pipeline is a more feasible and realistic approach \eg using LLMs only when their general-purpose contextual reasoning provides the most benefit and to navigate semantically challenging cases to improve recall, while relying on traditional, deterministic and specialized approaches for more straight-forward, rule-based analysis.

\section{Threats to Validity and Limitations}
\label{sec:limitation}

The main objective of this study is to analyze whether off-the-shelf LLMs are capable of replicating different facets of privacy policy analysis, both the end-to-end functionalities as well as the intermediate tasks. Hence, our study design has a number of limitations that we address accordingly, as we describe below. 

\myparagraph{Repeating experiments with System Prompts} 
To address concerns that LLM outputs may vary across repeated experiments simply due to how the prompt is formulated \ie a single prompt with all details in it, vs using a system prompt shared across all privacy policies paired with separate user prompts for each privacy policy input. 
In the latter, the LLM is configured to have a static system prompt across all analyses that describes the task definition, analysis guidelines, and the expected output format, while each query supplies the user prompt that provides the input privacy policy along with any specific analysis request (an example of which is in \secref{app:policylint_system_prompts}).
Note that the formulation of a system prompt is supported by the APIs of both the LLMs.  

We repeated our analysis in Section~\ref{contradiction-detection} \ie for \funcnumber{1}, and our results did not change substantially. 
With system prompt enabled, the number of contradictions detected increased by 4 (\ie 124 vs 120). 
However, the precision scores remained in a comparable range (\tabref{tab:policylint_system_prompt}): Gemini-2.5-Pro-\promptnum{2} still achieved the highest precision at 25.0\% (matching \tabref{tab:perf_overview}), while GPT-5.2-\promptnum{2} decreased from 26.3\% to 20.8\%. This suggests that simply changing the prompting structure does not have too much of an impact on our underlying findings.




\myparagraph{Additional Prompting Strategies} 
LLMs have been known to be highly sensitive to prompting strategies~\cite{prompt-sensitivity}. To probe whether additional prompting strategies might change our results significantly, we test our results against an additional prompting strategy 
by adopting an observation from LLM literature, 
\ie we improve on \promptnum{2} to create \promptnum{2A}, which includes example input-output pair that matches the corresponding tool's result format. To not increase the token count and thus incurring significant cost, we limit the provided example to one for each prompt. 
The results of this experiment are in \tabref{tab:policylint_prompt_2a}. With \promptnum{2A}, we did not see consistent or substantial improvements over the other two prompting strategies (\eg GPT-5.2-\promptnum{2A} reached 23.3\% precision vs.\ 26.3\% for \promptnum{2} in \tabref{tab:perf_overview}), and the results were consistent with some of the variance we observed between \promptnum{1} and \promptnum{2}. This suggests that further improvement in prompt quality may not yield predictable or consistent results.


\myparagraph{Grounding LLM Outputs in Source Policies} 
An additional concern when evaluating LLM outputs is whether the produced output is a result of analyzing the policy text provided as input, or if it is a {\em plausible-looking} result that cannot be traced back to the input policy text \ie {\em did the model genuinely analyzed the input policy text and produce the corresponding output?} 
This analysis was feasible since we instructed LLMs to provide the source sentence from which the result for any functionality is produced, as we mentioned in Section~\ref{methodology-pipeline}. 
To perform this analysis, we collect all source sentences provided as evidence by all LLM-prompt configurations (including \promptnum{2A} mentioned above) for \funcnumber{1} (\ie the PolicyLint tool). We chose this functionality due to its subjective nature, which might cause LLMs to make mistakes. 
Out of the total 1504 sentences in this analysis, we considered a sentence grounded if i) we found an exact matching sentence in the input policy text after text normalization (\ie lowercase, punctuation removal), or ii) if we found a sentence with at least a 90\% match when performing a fuzzy string-matching. 
Note that we chose a naive (\ie exact) or a fuzzy-based matching as opposed to more advanced similarity matching (\eg cosine-similarity) since the sentences were expected to be {\em verbatim} (or a very close match) from the policy text. 

While none of the LLM-prompt combination had a 100\% match, all scored over 90\%, with the least match of 92.2\% for GPT-5.2-\promptnum{2}, and the highest match of 98.2\% for Gemini-2.5-pro-\promptnum{2}. 
This shows that the LLM results were overwhelmingly grounded in the input policy text. Note that ``ungrounded'' here does not indicate an LLM hallucination~\cite{llm-hallunication}, \ie when we manually evaluated such cases, they could be attributed to sentence truncation (\eg 
{\em ``We do not knowingly collect... biometric information...''}) or merging multiple sentences rather than producing a fabricated sentence. 

\myparagraph{Privacy Policy Dataset Size} 
Given the cost associated with this study (see Section~\ref{sec:cost-analysis}), we deliberately limited our analysis to ten privacy policies, although prior privacy policy analysis work may have larger dataset~\cite{large-pp}. 
Instead, to broaden representation and impact, we choose the privacy policies of the top-ten most popular apps in the Google Play Store at the time of our experiments, as we discuss in Section~\ref{methodology-pipeline}. 
However, we acknowledge that the findings in this study may not generalize to other privacy policies, or other analyses done over a larger dataset, though we argue that the latter may not be feasible due to additional cost (Section~\ref{sec:cost-analysis}).

\section{Related Work}
\label{sec:related_work}

\myparagraph{LLM and Privacy Policy Research}
Privacy policy analysis research is increasingly leveraging LLMs to help users understand or analyze privacy policies, along various dimensions~\cite{llm1, llm2, llm3, llm4}. For instance, Sun~\etal develop an LLM-based agent to act as a privacy policy expert to help users understand website privacy policies~\cite{sun2025}, while Freiberger~\etal develop a browser extension that leverages LLM to help users understand website privacy policies~\cite{freiberger2025}. Similarly, Zhang~\etal propose PrivCAP that aims to enhance user comprehension of privacy policies by making it concise and interactive~\cite{zhang2025privcaptcha}. 
Rodriguez~\etal develop various guidance on the optimal design of prompts, parameters, and models, to leverage them for analyzing privacy policies at scale. 
Our work is fundamentally different,
 as we compare the extent to which an off-the-shelf LLM can subsume the tasks performed previously by specialized tools to establish a baseline, and hence, we deliberately assume multiple prompting strategies mimicking multiple stakeholders to establish the expertise needed to leverage LLMs in performing these tasks, rather than being guided by the best practices established by Rodriguez~\etal~\cite{rodriguez2024large}.  

\myparagraph{Privacy Policy Research Pre-LLM}
Before LLM, prior work primarily relied on various NLP techniques to perform privacy policy analysis work, and has developed tools and frameworks that cover a wide array of functionalities, and various aspects of research on privacy policies such as their availability in mobile apps~\cite{Bowers:SOUPS2017, Zimmeck:NDSS2017, Han:PETS2020, Degeling:NDSS2018}, 
readability and comprehension~\cite{McDonald:ISJLP2008, Bowers:WISEC2019, Jensen:CHI2004}, 
as well as analyzing their vagueness~\cite{amw20, Bhatia:RE2016,hfl18}, 
consent and opt-out choices~\cite{Sathyendra:ACL2017, Okoyomon:CONPRO2019}, 
contradictions~\cite{Andow:SECURITY2019, Cranor:TWEB2016, Yu:DSN2016, hfl18}, 
and regulatory compliance~\cite{Bowers:SOUPS2017, Bowers:WISEC2019, manandhar-sec22}.
Our research is fundamentally different as it instead performs a comparative study of LLMs against this body of work, and evaluates to what extent LLMs are capable of performing the functionalities and intermediate methodology tasks of the tools proposed in these prior work.


\section{Conclusion}
\label{sec:conclusion}

In this paper, we conducted the first systematic evaluation of whether off-the-shelf LLMs can replace specialized privacy policy and compliance analysis tools, spanning three functionalities (contradiction detection, regulatory compliance analysis, and policy aggregation), two intermediate tasks (semantic role labeling and human manual annotation). 
Across nearly all tasks, off-the-shelf LLMs, prompted with varying levels of detail but without any domain-specific fine-tuning, matched or exceeded the performance of specialized tools. As our findings (\fnumber{1}$\rightarrow$\fnumber{15}) show, LLMs can perform desired privacy policy analysis functionality even when receiving only a high-level instruction, while not being bound by limitations that existed in prior tools.
Taken together, our findings suggest that the privacy policy analysis landscape is at an inflection point: while LLMs show the ability to subsume many functionalities and methodologies of prior research allowing for the development of even more sophisticated tools in this domain, they also simultaneously introduce significantly higher monetary costs and novel design challenges that did not exist in prior tools, indicating that a hybrid approach may be preferable, that balances the flexibility of LLMs with the scalability from prior tools.


\bibliographystyle{plainurl}
\bibliography{bibliographies/os,bibliographies/policy,bibliographies/iot,bibliographies/misc,bibliographies/politics,bibliographies/privacy,bibliographies/llms}

\appendix

\begin{table*}[t]
\centering
\caption{LLM performance overview per task. Metrics are percentages. Tasks marked with * lack ground truth and report Precision (P) and agreement count (N) only. Purpliance and PolicyPulse report both the 10-policy stratified sample and the TikTok full-policy evaluation.}
\label{tab:perf_overview}
\footnotesize
\setlength{\tabcolsep}{6pt}
\renewcommand{\arraystretch}{1.14}
\setlength{\extrarowheight}{2pt}
\begin{tabular}{@{}l l c c c c c@{}}
\hline
 & & \multicolumn{2}{c}{\textbf{Gemini-2.5-Pro}} & \multicolumn{2}{c}{\textbf{GPT-5.2}} & \\
\cline{3-6}
\textbf{Task} & \textbf{Metric} & \promptnum{1} & \promptnum{2} & \promptnum{1} & \promptnum{2} & \textbf{Tool} \\
\hline
AutoCompliance (GDPR Completeness) & P & 74.0 & 88.0 & \textbf{89.0} & \textbf{89.0} & 59.0 \\
 & A & 91.0 & 95.0 & \textbf{97.0} & 96.0 & 78.0 \\
 & R & \textbf{100.0} & 92.0 & \textbf{100.0} & 96.0 & 40.0 \\
 & F1 & 85.0 & 90.0 & \textbf{94.0} & 92.0 & 48.0 \\
\hline
PolicyChecker (GDPR Completeness) & P & 77.8 & 92.9 & 67.0 & \textbf{94.3} & 73.3 \\
 & A & 47.7 & \textbf{73.1} & 26.2 & 72.5 & 37.7 \\
 & R & 89.0 & \textbf{92.9} & 83.8 & 91.8 & 83.9 \\
 & F1 & 79.1 & \textbf{89.9} & 60.4 & 89.4 & 68.7 \\
\hline
Poligraph (Policies Summarization) & P & 93.4 & 91.7 & 90.7 & 91.8 & \textbf{98.2} \\
 & A & 78.6 & 71.4 & 78.6 & \textbf{81.0} & 57.9 \\
 & R & 80.2 & 72.6 & 83.0 & \textbf{84.9} & 50.9 \\
 & F1 & 86.3 & 81.0 & 86.7 & \textbf{88.2} & 67.1 \\
\hline
Poligraph* (CCPA Terms Detection) & P & \textbf{76.6} & 61.9 & 53.0 & 52.3 & 66.7 \\
 & N & \textbf{49/64} & 39/63 & 79/149 & 78/149 & 4/6 \\
\hline
PolicyLint* (Contradiction Detection) & P & 17.1 & 25.0 & 9.1 & \textbf{26.3} & 0.0 \\
 & N & 6/35 & 7/28 & 4/44 & \textbf{5/19} & 0/5 \\
\hline
Purpliance* (10-policy sample) & P & 79.9 & \textbf{96.2} & 76.7 & 76.7 & 79.8 \\
 & N & 131/164 & \textbf{154/160} & 92/120 & 132/172 & 99/124 \\
\hline
Purpliance* (TikTok full policy) & P & 91.1 & \textbf{98.7} & 86.1 & 98.1 & 83.8 \\
 & N & 448/492 & \textbf{233/236} & 458/532 & 369/376 & 352/420 \\
\hline
PolicyPulse* (10-policy sample) & P & 71.7 & 73.9 & \textbf{88.1} & 65.0 & 55.4 \\
 & N & 132/184 & 204/276 & \textbf{733/832} & 156/240 & 286/516 \\
\hline
PolicyPulse* (TikTok full policy) & P & 70.1 & 82.4 & \textbf{91.1} & 72.3 & 59.9 \\
 & N & 101/144 & 234/284 & \textbf{889/976} & 269/372 & 405/676 \\
\hline
\end{tabular}
\end{table*}

\subsection{Semantic Role Labeling Data Class Precision}
\label{app:srl_field_precision}

\begin{figure}[htbp]
    \centering
    \includegraphics[width=\columnwidth]{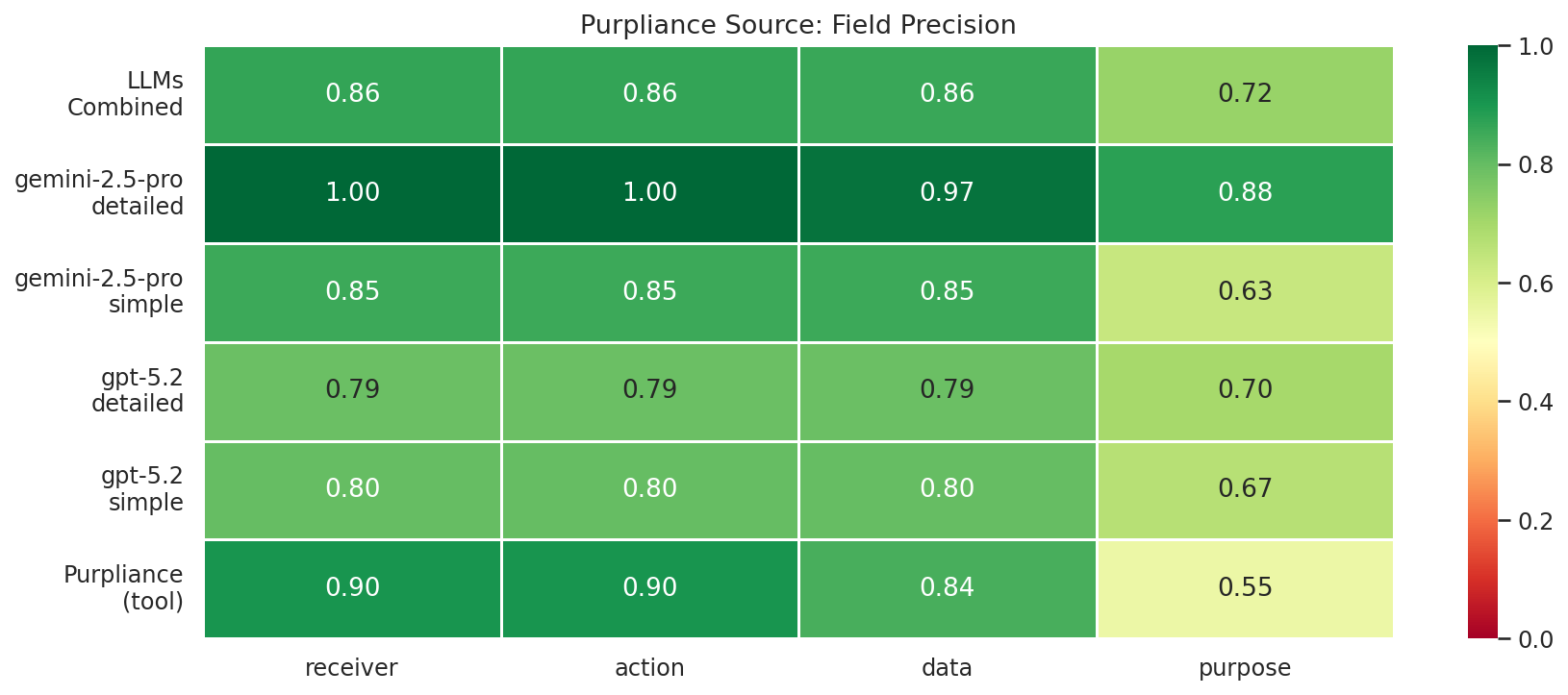}
    \caption{Data Class Level Precision: Purpliance vs.\ LLMs.}
    \label{fig:purpliance_field_precision_heatmap}
\end{figure}

\begin{figure}[htbp]
    \centering
    \includegraphics[width=\columnwidth]{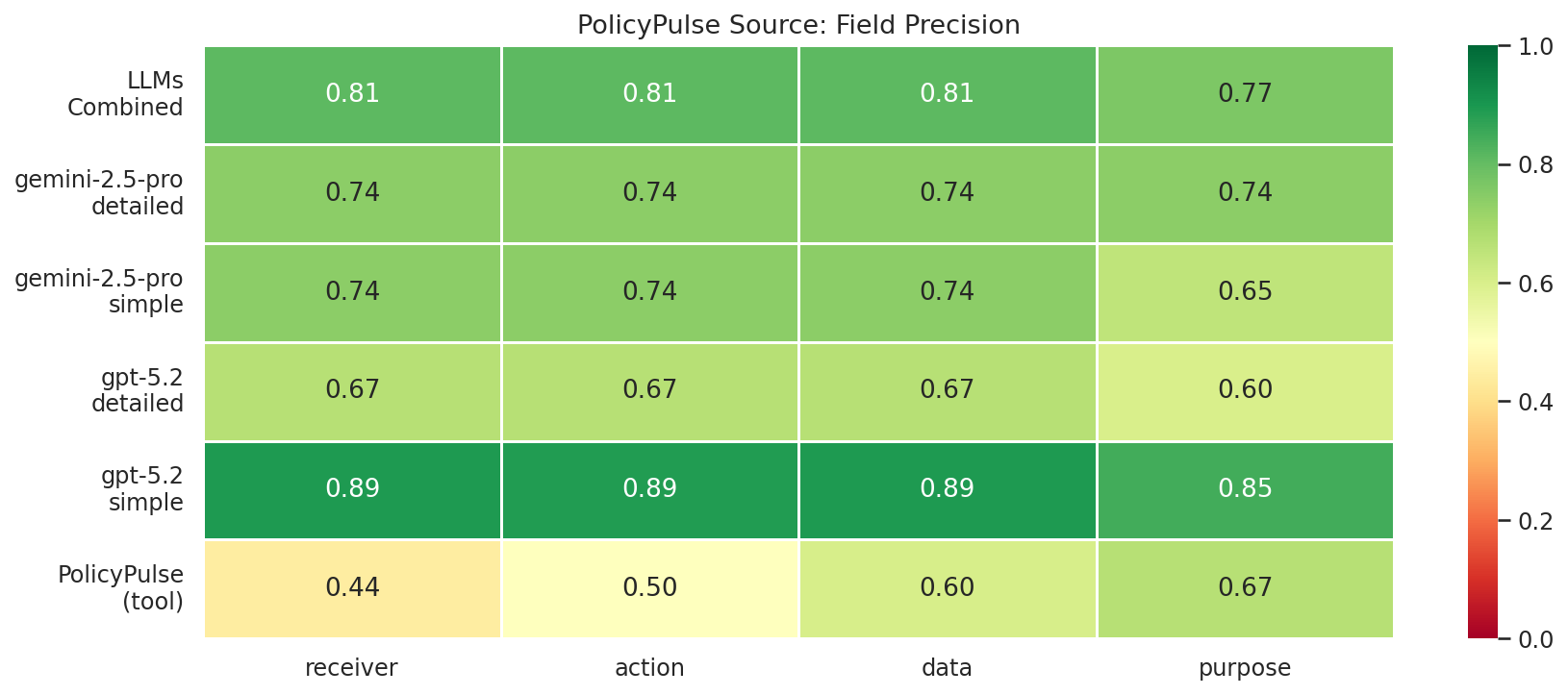}
    \caption{Data Class Level Precision: PolicyPulse vs.\ LLMs.}
    \label{fig:policypulse_field_precision_heatmap}
\end{figure}

Data class level precision heatmaps for Purpliance and PolicyPulse across the four scored data classes (\texttt{data}, \texttt{action}, \texttt{receiver}, and \texttt{purpose}), comparing each tool with LLM configurations under thinking mode. Figure \ref{fig:purpliance_field_precision_heatmap} and \ref{fig:policypulse_field_precision_heatmap} shows the data class level precision for Purpliance and PolicyPulse respectively.

\subsection{Detailed(\promptnum{2}) LLM Example}
\label{app:detailed_prompts}

\lstdefinestyle{appendixprompt}{
  language=,
  basicstyle=\scriptsize\ttfamily,
  breaklines=true,
  breakatwhitespace=false,
  breakindent=0pt,
  columns=fullflexible,
  keepspaces=true,
  showstringspaces=false,
  frame=none,
  aboveskip=0pt,
  belowskip=0pt,
}
\newcommand{\promptbox}[1]{%
  \begin{mdframed}[
    linewidth=0.4pt,
    innerleftmargin=3pt,
    innerrightmargin=3pt,
    innertopmargin=3pt,
    innerbottommargin=3pt,
    skipabove=4pt,
    skipbelow=4pt,
  ]
  \lstinputlisting[style=appendixprompt]{#1}%
  \end{mdframed}%
}

This appendix lists an {\em example} of the prompts used to instruct GPT-5.2 and Gemini-2.5-Pro in each functionality replication task (\secref{sec:func-replication}). In all prompts, \texttt{<INPUT>} is replaced at runtime with the plain-text privacy policy under analysis. We provide \promptnum{2}, \ie detailed prompt, for the contradiction analysis task for PolicyLint as an example.

\label{app:prompt_contradiction_policylint}
\promptbox{appendix/prompts/contradiction_policylint_detailed.txt}
\subsection{System and User Prompt Example (PolicyLint)}
\label{app:policylint_system_prompts}

The system-prompt experiment (see Section~\ref{sec:limitation}) splits the combined \promptnum{2} prompt into a fixed \emph{system} message (role and JSON output schema) and a per-policy \emph{user} message (contradiction criteria plus \texttt{<INPUT>}).

\textbf{System prompt}
\promptbox{appendix/prompts/contradiction_policylint_system_prompt.txt}

\textbf{User prompt (\promptnum{2})}
\promptbox{appendix/prompts/contradiction_policylint_user_detailed.txt}

\begin{table}[t]
\centering
\caption{OPP-115 category-level metrics vs.\ consolidated (1.0) ground truth. Micro-averaged Precision (P), Recall (R), and F1 (\%). 
}
\label{tab:opp115_category_metrics}
\footnotesize
\setlength{\tabcolsep}{3.5pt}
\renewcommand{\arraystretch}{1.1}
\begin{tabular}{@{}l l c c c c@{}}
\hline
 & & \multicolumn{2}{c}{\textbf{Gemini-2.5-Pro}} & \multicolumn{2}{c}{\textbf{GPT-5.2}} \\
\cline{3-6}
\textbf{Category} & \textbf{Metric} & \promptnum{1} & \promptnum{2} & \promptnum{1} & \promptnum{2} \\
\hline
\multirow{3}{*}{\makecell[l]{Cat.~A\\1st party}} & P & \textbf{76.6} & 66.7 & 74.0 & 74.4 \\
 & R & 72.4 & \textbf{80.0} & 62.8 & 60.0 \\
 & F1 & \textbf{74.5} & 72.7 & 67.9 & 66.4 \\
\hline
\multirow{3}{*}{\makecell[l]{Cat.~B\\3rd party}} & P & 87.7 & 85.5 & 89.9 & \textbf{90.7} \\
 & R & 82.3 & 81.5 & 61.5 & \textbf{75.4} \\
 & F1 & \textbf{84.9} & 83.5 & 73.1 & 82.4 \\
\hline
\end{tabular}
\end{table}

\begin{table}[t]
\centering
\caption{OPP-115 per-human annotator comparison (thinking mode, categories 1 and 3 combined). F1 (\%) of each LLM configuration against the individual human annotator's labels on that annotator's policy subset. 
}
\label{tab:opp115_per_human}
\footnotesize
\setlength{\tabcolsep}{3.5pt}
\renewcommand{\arraystretch}{1.1}
\begin{tabular}{@{}l c c c c@{}}
\hline
 & \multicolumn{2}{c}{\textbf{Gemini-2.5-Pro}} & \multicolumn{2}{c}{\textbf{GPT-5.2}} \\
\cline{2-5}
\textbf{Human} & \promptnum{1} & \promptnum{2} & \promptnum{1} & \promptnum{2} \\
\hline
\#82  & \textbf{80.3} & 75.9 & 70.6 & 78.0 \\
\#84  & 75.6 & 74.0 & 75.2 & 70.8 \\
\#88  & \textbf{78.8} & 77.3 & 73.7 & 76.3 \\
\#95  & \textbf{78.0} & 69.8 & 68.2 & 74.4 \\
\#103 & \textbf{78.0} & 69.8 & 72.7 & 74.4 \\
\#116 & \textbf{71.1} & 69.8 & 63.9 & 66.7 \\
\#117 & \textbf{73.7} & 69.3 & 62.8 & 64.9 \\
\#118 & 74.6 & 70.1 & 68.9 & \textbf{74.7} \\
\#121 & 66.1 & 62.9 & 68.0 & \textbf{69.9} \\
\hline
\textbf{Average} & \textbf{75.1} & 71.0 & 69.3 & 72.2 \\
\hline
\end{tabular}
\end{table}

\begin{table}[t]
\centering
\caption{OPP-115 category annotation vs.\ consolidated (1.0) ground truth. Segment-level Precision (P), Accuracy (A), Recall (R), and F1 reported as percentages across categories 1 and 3. Default prompts omit category definitions; \textit{with definitions} includes them.}
\label{tab:perf_opp115_thinking}
\footnotesize
\setlength{\tabcolsep}{4pt}
\renewcommand{\arraystretch}{1.1}
\begin{tabular}{lcccc}
\hline
 & \multicolumn{2}{c}{\textbf{Gemini-2.5-Pro}} & \multicolumn{2}{c}{\textbf{GPT-5.2}} \\
\cline{2-5}
\textbf{Metric} & \promptnum{1} & \promptnum{2} & \promptnum{1} & \promptnum{2} \\
\hline
P & \textbf{81.8} & 74.5 & 80.7 & 82.2 \\
A & \textbf{85.4} & 82.8 & 80.7 & 82.7 \\
R & 77.1 & \textbf{80.7} & 62.2 & 67.3 \\
F1 & \textbf{79.4} & 77.5 & 70.2 & 74.0 \\
\hline
\end{tabular}g
\end{table}

\begin{table}[t]
\centering
\caption{PolicyLint contradiction detection with system prompting. Precision (P) and agreement count (N) from human evaluation.}
\label{tab:policylint_system_prompt}
\footnotesize
\setlength{\tabcolsep}{4pt}
\renewcommand{\arraystretch}{1.1}
\begin{tabular}{@{}l c c c c@{}}
\hline
 & \multicolumn{2}{c}{\textbf{Gemini-2.5-Pro}} & \multicolumn{2}{c}{\textbf{GPT-5.2}} \\
\cline{2-5}
\textbf{Metric} & \promptnum{1} & \promptnum{2} & \promptnum{1} & \promptnum{2} \\
\hline
P & 23.1 & \textbf{25.0} & 19.6 & 20.8 \\
N & 9/39 & \textbf{6/24} & 10/51 & 5/24 \\
\hline
\end{tabular}
\end{table}

\begin{table}[t]
\centering
\caption{PolicyLint contradiction detection with \promptnum{2A}. \promptnum{2A} augments \promptnum{2} with one example input--output pair matching the tool result format. Precision (P) and agreement count (N) from human evaluation.}
\label{tab:policylint_prompt_2a}
\footnotesize
\setlength{\tabcolsep}{4pt}
\renewcommand{\arraystretch}{1.1}
\begin{tabular}{@{}l c c@{}}
\hline
\textbf{Metric} & \textbf{Gemini-2.5-Pro} & \textbf{GPT-5.2} \\
\hline
P & 23.1 & \textbf{23.3} \\
N & 6/26 & \textbf{10/43} \\
\hline
\end{tabular}
\end{table}

\end{document}